\documentclass[prc,superscriptaddress,unsortedaddress,twocolumn,showpacs]{revtex4}
\usepackage{graphicx}
\usepackage{amsmath}
\usepackage{amssymb}
\usepackage{times}
\usepackage{bm}
\usepackage{braket}

\usepackage{color}
\usepackage{ulem}

\def\la{\mathrel{\mathpalette\fun <}}
\def\ga{\mathrel{\mathpalette\fun >}}
\def\fun#1#2{\lower3.6pt\vbox{\baselineskip0pt\lineskip.9pt
\ialign{$\mathsurround=0pt#1\hfil##\hfil$\crcr#2\crcr\sim\crcr}}}

\begin{document}

\title{
Investigation of $\alpha$ clustering with knockout reactions
}

\author{Kazuki Yoshida}
\email[]{yoshida.kazuki@jaea.go.jp}
\affiliation{Advanced Science Research Center, Japan Atomic Energy Agency, 
Tokai, Ibaraki 319-1195, Japan}
\affiliation{Research Center for Nuclear Physics (RCNP), Osaka
University, Ibaraki 567-0047, Japan}

\author{Kazuyuki Ogata}
\affiliation{Research Center for Nuclear Physics (RCNP), Osaka
University, Ibaraki 567-0047, Japan}
\affiliation{Department of Physics, Osaka City University, Osaka 558-8585, Japan}

\author{Yoshiko Kanada-En'yo}
\affiliation{Department of Physics, Kyoto University, Kyoto 606-8502, Japan}

\date{\today}

\begin{abstract}
\begin{description}
\item[Background]
Nuclear clustering has been one of the main interests in nuclear physics.
In order to probe the $\alpha$ clustering through reaction observables,
$\alpha$ transfer and $\alpha$ knockout reactions have been studied.
It is very important to probe the $\alpha$ cluster amplitude at nuclear
surface since the $\alpha$ spectroscopic factor is not necessarily a direct
measure of the $\alpha$ clustering.
\item[Purpose]
Our goal is to reveal how the
$\alpha$ cluster amplitude is probed through $\alpha$
knockout reactions depending on reaction conditions, e.g.,
the incident energy.
\item[Method]
We consider $^{20}$Ne($p$, $p\alpha$)$^{16}$O and $^{120}$Sn($p$, $p\alpha$)$^{116}$Cd
at 100--400~MeV
within the distorted wave impulse
approximation (DWIA) framework.
We introduce a masking function, which shows how
the reaction amplitude in the nuclear
interior is suppressed and defines the probed region of
the $\alpha$ cluster
wave function.
\item[Results]
It is clearly shown by means of the masking function
that the $\alpha$ knockout reaction probes
the $\alpha$ cluster amplitude in the nuclear surface region,
which is the direct measure of well-developed $\alpha$ cluster states.
The incident energy dependence of the masking effect is
investigated, using a simplified form of the masking function.
\item[Conclusions]
The $\alpha$ knockout reaction can probe the $\alpha$ cluster amplitude
in the nuclear surface region by choosing proper kinematics
owing to the masking effect originated
from absorptions of distorting potentials, and is a suitable method
to investigate how $\alpha$ cluster states are spatially developed.

\end{description}
\end{abstract}


\maketitle

\section{Introduction}
Cluster states are one of the main interests in nuclear physics.
For a recent review, see Ref.~\cite{Hor12}.
When investigating $\alpha$ cluster states,
it should be noted that
a large $\alpha$ spectroscopic factor
does not necessarily indicate
well-developed $\alpha$ cluster states
because of the dual nature of the  cluster and
the shell model~\cite{Bay58}.
From this point of view,
recently the $^{16}$O($^{6}$Li,$d$)$^{20}$Ne
$\alpha$ transfer reaction has been studied~\cite{Fuk16}
with a three-body reaction model using a macroscopic
cluster wave function,
and the reaction was shown to have high sensitivity
in the nuclear surface region and suitable
to prove the $\alpha$ cluster amplitude there,
i.e., spatially developed $\alpha$ cluster states.

An alternative method to the $\alpha$ transfer reaction is the
proton-induced $\alpha$ knockout reaction, i.e., ($p$, $p\alpha$).
In the present study we employ the distorted wave impulse approximation (DWIA)
framework, which has been utilized and well established in $\alpha$ knockout
reaction studies~\cite{Car84,Mab09,Roo77,Nad80,Nad89,Wan85,Yos88}
and nucleon knockout reactions~\cite{Sam86,Cha77,Sam87,Jac66,Jac73,Kit85}
as well.
For a recent review on the ($p$, $p$N) reactions,
see Ref.~\cite{Wak17}.
In this paper we examine the
peripherality of $^{20}$Ne($p$, $p\alpha$)$^{16}$O and
its incident energy dependence to investigate how the $\alpha$ cluster
amplitude in the nuclear surface region is probed through
the $^{20}$Ne($p$, $p\alpha$)$^{16}$O reaction.
The $^{120}$Sn($p$, $p\alpha$)$^{116}$Cd reaction is also investigated
in a similar manner.

In Sec.~\ref{sectheory} we describe the DWIA formalism for
($p$, $p\alpha$) reactions and also how the cluster wave function is
constructed in the present study.
The definition of the masking function,
which is the key concept in the present study,
is also given.
In Sec.~\ref{secresult} we introduce a general feature of
knockout reactions.
Then the absorption effect in $\alpha$ knockout reactions
due to the distorting potential is discussed in terms of the
masking function, which defines the probed region in the ($p$, $p\alpha$)
reaction.
The $^{120}$Sn($p$, $p\alpha$)$^{116}$Cd reaction is
investigated as a case of strong
absorption and Coulomb effects.
The incident energy dependence of the masking effect
is also discussed with introducing a simplified form of
the masking function.
Finally, a summary is given in Sec.~\ref{secsum}.

\section{Theoretical framework}
\label{sectheory}

\subsection{DWIA formalism}
\label{subsecDWIA}
In the present study we consider the $^{20}$Ne($p$, $p\alpha$)$^{16}$O reaction
in normal kinematics.
The incoming and outgoing protons are labeled as particle
0 and 1, respectively.
${\bm K}_i$, $\Omega_i$, $E_i$, and $T_i$ denote
the momentum (wave number), its solid angle,
total and kinetic energy of particle $i$ ($=0, 1, \alpha$),
respectively.
All quantities appear below are evaluated in the
center-of-mass (c.m.) frame, except those with
the superscript L which are in the laboratory (L) frame.

In the DWIA framework, the transition amplitude of
A($p$, $p\alpha$)B is given by
\begin{align}
T_{{\bm K}_i}^{nlm}
&=
\Braket{
\chi_{1,{\bm K}_1}^{(-)}
\chi_{\alpha,{\bm K}_\alpha}^{(-)}
|t_{p\alpha}|
\chi_{0,{\bm K}_0}^{(+)}
\varphi_{\alpha}^{nlm}
},
\label{eqtmt}
\end{align}
where $\chi_{i,{\bm K}_i}$ ($i=0, 1, \alpha$) are the
distorted waves of $p$-A, $p$-B, and $\alpha$-B systems with
relative momentum (wave number) ${\bm K}_i$,
respectively.
The superscripts $(+)$ and $(-)$ are given to specify the outgoing and
the incoming boundary conditions of the scattering waves, respectively.
The $\alpha$ cluster wave function of the $\alpha$-B system is denoted by
$\varphi_{\alpha}^{nlm}$ with $n$, $l$, and $m$ being
the principal quantum number, the angular momentum, and its third component, respectively.
$t_{p\alpha}$ is the
effective interaction between $p$ and $\alpha$,
which is the transition interaction in the DWIA framework.

Applying the so-called factorization approximation,
or  the asymptotic momentum approximation, which has
been justified in Ref.~\cite{Yos16},
Eq.~(\ref{eqtmt}) is reduced to
\begin{align}
T_{{\bm K}_i}^{nlm}
&\approx
\tilde{t}_{p\alpha}({\bm \kappa}',{\bm \kappa})
\int d{\bm R}\,
F_{{\bm K}_i}({\bm R})\,
\varphi_{\alpha}^{nlm}({\bm R}),
\label{eqfact}
\end{align}
where ${\bm \kappa}$ (${\bm \kappa}'$) is the $p$-$\alpha$
relative momentum in the initial (final) state.
$\tilde{t}_{p\alpha}$ and $F_{{\bm K}_i} ({\bm R})$ are defined by
\begin{align}
\tilde{t}_{p\alpha}({\bm \kappa}',{\bm \kappa})
&\equiv
\int d{\bm s}\,
e^{-i{\bm \kappa}'\cdot{\bm s}}\,
t_{p\alpha}({\bm s})\,
e^{i{\bm \kappa}\cdot{\bm s}}, \\
F_{{\bm K}_i}({\bm R})
&\equiv
\chi_{1,{\bm K}_1}^{*(-)}({\bm R})
\chi_{\alpha,{\bm K}_\alpha}^{*(-)}({\bm R})
\chi_{0,{\bm K}_0}^{(+)}({\bm R})\,
e^{-i{\bm K}_0\cdot{\bm R}A_{\alpha}/A},
\label{eqF}
\end{align}
where $A_\alpha = 4$ and $A$ is the mass number of the nucleus A.
By making the on-the-energy-shell (on-shell) approximation
of the final state prescription:
\begin{align}
\bm{\kappa}
&=
\kappa'\hat{\bm{\kappa}},
\end{align}
the matrix element of $t_{p\alpha}$ in Eq.~(\ref{eqfact})
can be related with a free $p$-$\alpha$ differential cross section as
\begin{align}
\left|
\tilde{t}_{p\alpha}({\bm \kappa}',{\bm \kappa})
\right|^2
\approx
\frac{(2\pi\hbar ^2)^2}{\mu_{p\alpha}^2}
\frac{d\sigma_{p\alpha}}{d\Omega_{p\alpha}}(\theta_{p\alpha},T_{p\alpha}),
\end{align}
where the $p$-$\alpha$ scattering angle $\theta_{p\alpha}$
is an angle between ${\bm \kappa}$ and ${\bm \kappa}'$,
and the scattering energy is defined by
$T_{p\alpha} = (\hbar\kappa')^{2} / (2\mu_{p\alpha})$ with
$\mu_{p\alpha}$ being the reduced mass of the $p$-$\alpha$ system.
The triple differential cross section (TDX) of the
A($p$, $p\alpha$)B reaction is then given by
\begin{align}
\frac{d^3\sigma}{dE_1^{\mathrm L}d\Omega_1^{\mathrm L} \Omega_2^{\mathrm L}}
&=
S_{\alpha}F_{\mathrm{kin}} C_0
\frac{d\sigma_{p\alpha}}{d\Omega_{p\alpha}}(\theta_{p\alpha},T_{p\alpha})
\sum_{m}
\left|
\bar{T}_{{\bm K}_i}^{nlm}
\right|^2,
\end{align}
where $S_\alpha$ is the so-called $\alpha$ spectroscopic factor, and
$F_{\mathrm{kin}}$, $C_0$, and
$\bar{T}_{{\bm K}_i}^{nlm}$ are defined by
\begin{align}
F_{\mathrm{kin}}
&\equiv
J_{\mathrm L}\frac{K_1 K_\alpha E_1 E_\alpha}{(\hbar c)^4}
\left[
1+\frac{E_\alpha}{E_B}+\frac{E_\alpha}{E_B}\frac{{\bm K}_1\cdot{\bm K}_\alpha}{K_\alpha^2}
\right]^{-1}, \\
C_0
&\equiv
\frac{E_0}{(\hbar c)^2 K_0}
\frac{1}{2l+1}
\frac{\hbar ^4}{(2\pi)^3 \mu_{p\alpha}^2},
\end{align}
and
\begin{align}
\bar{T}_{{\bm K}_i}^{nlm}
&\equiv
\int d{\bm R}\,
F_{{\bm K}_i}({\bm R})\,
\varphi_{\alpha}^{nlm}({\bm R})
\label{eqtbar}
\end{align}
with $J_{\mathrm L}$ being the Jacobian from the c.m. frame to the L frame.

The DWIA framework introduced in this section
has been validated in Ref.~\cite{Yos18} by the benchmark comparison with
the transfer-to-the-continuum model \cite{Mor15,Mar17}
and the Faddeev/Alt-Grassberger-Sandhas method \cite{Fad61,Alt67}.

\subsection{$\alpha$ cluster wave function}
The $\alpha$ cluster wave function $\varphi_{\alpha}^{nlm}({\bm R})$
for the $\alpha$-$^{16}$O system
is defined as the eigenstate of the Schr\"odinger equation:
\begin{align}
\left[
-\frac{\hbar^2}{2\mu_{\alpha}}\nabla_{\bm R}^2
+V_{\alpha \mathrm{B}}(R)
\right]\varphi_{\alpha}^{nlm}({\bm R})
&=
\varepsilon_{\alpha}\varphi_{\alpha}^{nlm}({\bm R}),
\label{eqsch-alpha}
\end{align}
where $\mu_\alpha$, $V_{\alpha \textrm{B}}$, and
$\varepsilon_{\alpha}$ are
the reduced mass, the binding potential, and the binding energy
of $\alpha$-B system, respectively.

For the nuclear part of $V_{\alpha \mathrm{B}}$,
a central Woods-Saxon shape with
the radius parameter $r_0$ and the diffuseness parameter $a_0$
is adopted:
\begin{align}
f_{\mathrm{WS}}(R)
&=
\frac{1}{1+\mathrm{exp}\left(\displaystyle\frac{R-r_0}{a_0}\right)}.
\end{align}
The numerical input for $r_0$, $a_0$ and $\varepsilon_{\alpha}$
are given in Sec.~\ref{subsecnum}.
Since $V_{\alpha \mathrm{B}}$ is a central potential,
$\varphi_{\alpha}^{nlm}({\bm R})$ can be simply separated into the radial
and the angular part as
\begin{align}
\varphi_{\alpha}^{nlm}({\bm R})
=
\phi_{\alpha}^{nl}(R)Y_{lm}(\Omega),
\end{align}
where $\phi_{\alpha}^{nl}(R)$, $Y_{lm}$, and $\Omega$ are
the radial part of $\varphi_{\alpha}^{nlm}(\bm{R})$,
the spherical harmonics, and
the solid angle of ${\bm R}$, respectively.

\subsection{Masking function}
Since the square modulus of Eq.~(\ref{eqtbar})
is proportional to the knockout cross section,
it should be worth investigating
the property of the integrand in the right-hand side of Eq.~(\ref{eqtbar})
to reveal a contribution of $\varphi_{\alpha}^{nlm}$
to the knockout cross section.
Here we define the masking function
\begin{align}
D_{lm}(R)
\equiv
\frac{1}{\sqrt{4\pi}}
\int d\Omega\,
 F_{{\bm K}_i}({\bm R}) Y_{lm}(\Omega)
\end{align}
so that $\bar{T}_{{\bm K}_i}^{nlm}$ is given by
\begin{align}
\bar{T}_{{\bm K}_i}^{nlm}
&=
\sqrt{4\pi}
\int dR\,
R^2 D_{lm}(R)\, \phi_{\alpha}^{nl}(R).
\label{eqmask}
\end{align}
Therefore the masking function $D_{lm}(R)$ is a weighting function
which determines
the radial contribution of $\phi_{\alpha}^{nl}(R)$
to the knockout reaction amplitude.

If the nucleus B can be treated as a spectator,
the total momentum of the $p$-$\alpha$ system is approximately conserved;
\begin{align}
{\bm k}_{\alpha}
&\approx
\bm{q}
\equiv
{\bm K}_1+{\bm K}_{\alpha}
-\left(1-\frac{A_\alpha}{A}\right){\bm K}_0,
\label{eqmomtrans}
\end{align}
where $\bm{q}$ is the so-called missing momentum and
${\bm k}_{\alpha}$ is the momentum of the
$\alpha$ cluster in the nucleus A in the initial state.
It can be shown that $\bm{k}_\alpha$ is actually the momentum
$-\bm{K}_\mathrm{B}^\mathrm{L}$.
Once all distorting potentials are switched off, i.e.,
the plane wave impulse approximation (PWIA) is adopted,
$\bar{T}_{{\bm K}_i}^{nlm}$ turns out to be the Fourier transform
of the $\alpha$ cluster wave function,
\begin{align}
\bar{T}_{{\bm K}_i}^{nlm}
&\approx
\int d{\bm R}\,
e^{-i{\bm k}_{\alpha}\cdot{\bm R}}\phi_{\alpha}^{nl}({\bm R}).
\end{align}
The masking function $D_{lm}(R)$ is normalized to
unity when an $l=0$ cluster wave function is considered in PWIA,
as discussed in Sec.~\ref{subsecmask}.

\section{Results and discussion}
\label{secresult}
\subsection{Numerical inputs}
\label{subsecnum}
\begin{figure}[htbp]
\centering
\includegraphics[width=0.45\textwidth]{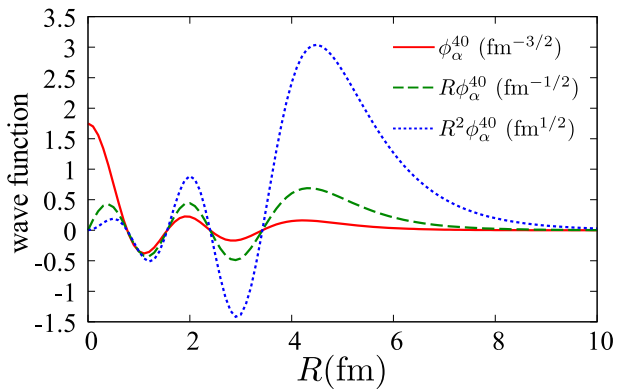}
\caption{
$\alpha$ cluster wave function obtained by solving
Eq.~(\ref{eqsch-alpha}).
The solid, dashed, and dotted lines correspond to
$\phi_\alpha^{40}(R)$, $R\phi_\alpha^{40}(R)$, and
$R^2\phi_\alpha^{40}(R)$, respectively.
The dashed line is equivalent to the PM1 in Fig.~3 of Ref.~\cite{Fuk16}.
}
\label{figcluster-w.f.}
\end{figure}

In the calculation of $^{20}$Ne($p$, $p\alpha$)$^{16}$O, $n=4$, and $l=0$ are assumed
for the $\alpha$ cluster orbital.
For the binding potential $V_{\alpha \mathrm{B}}(R)=V_0 f_{\mathrm{WS}}(R)$,
the radius parameter $r_0=1.25 \times 16^{1/3}$~fm and the
diffuseness parameter $a_0 = 0.76$~fm are employed.
These parameters are fixed to describe the behavior
of the microscopic cluster model wave function in the tail region~\cite{Fuk16}.
The depth parameter $V_0$ is determined so as to reproduced the
$\alpha$ separation energy of 4.73~MeV.
In Fig.~\ref{figcluster-w.f.} we show the obtained
$\phi_{\alpha}^{40}(R)$
by solving Eq.~(\ref{eqsch-alpha}).
The solid, dashed, and dotted lines are
$\phi_\alpha^{40}(R)$, $R\phi_\alpha^{40}(R)$, and
$R^2\phi_\alpha^{40}(R)$, respectively.
The dashed line is equivalent to the PM1 in Fig.~3 of Ref.~\cite{Fuk16}.
The dotted line
in Fig.~\ref{figcluster-w.f.}
shows a cluster wave function multiplied by $R^2$, which
will be helpful in the following discussion
because $R^2$ appears as a weight on $\phi_{\alpha}^{40}(R)$
in the transition matrix, as shown in Eq.~(\ref{eqmask}).

For the optical potentials of the incoming and outgoing protons,
the global optical potential by Koning and Delaroche~\cite{Kon03}
is employed, and for the $\alpha$-$^{16}$O potential in the final state,
the parameter set by Nolte, \textit{et al}.~\cite{Nol87} is employed.
For calculating the $p$-$\alpha$ scattering cross section,
the microscopic single folding model~\cite{Toy13} with a phenomenological
$\alpha$ density and the Melbourne nucleon-nucleon (NN) $g$-matrix interaction~\cite{Amo00}
is adopted.

\subsection{$^{20}$Ne($p$, $p\alpha$)$^{16}$O reaction}
\label{subsecmask}
\begin{figure}[htbp]
\centering
\includegraphics[width=0.45\textwidth]{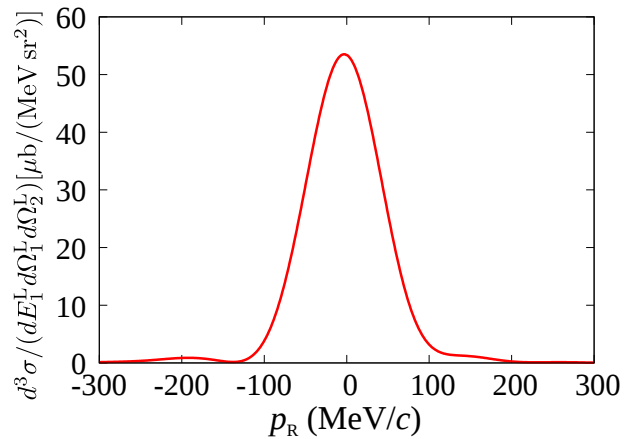}
\caption{
TDX of $^{20}$Ne($p$, $p\alpha$)$^{16}$O at 392~MeV as a function of
the recoil momentum.
}
\label{figDWvsPW}
\end{figure}

In Fig.~\ref{figDWvsPW} we show the TDX of $^{20}$Ne($p$, $p\alpha$)$^{16}$O at 392~MeV
as a function of the recoil momentum $p_{R}$ defined by
\begin{align}
p_{R}
&=
\hbar K_{\mathrm{B}}^{\mathrm{L}}
\frac{K_{\mathrm{B}z}^{\mathrm{L}}}{\left| K_{\mathrm{B}z}^{\mathrm{L}} \right|},
\end{align}
where $K_{\mathrm{B}z}^{\mathrm{L}}$ is the $z$-component of
$\bm{K}_{\mathrm{B}}^{\mathrm{L}}$ following the Madison convention.
The kinematics are fixed to be
$T_1=352$~MeV, $\theta_1^\mathrm{L} = 32.5^\circ$,
and $T_\alpha$ ($\theta_\alpha$) varies 31--35~MeV (27--108$^\circ$).
The azimuthal angles $\phi_{\bm{K}_1}$ and $\phi_{\bm{K}_2}$ are
fixed at 0 and $\pi$, respectively,
i.e., the scattered particles are in a coplanar.
The TDX has a peak at $p_{R}\sim 0$~MeV/$c$ when $l=0$, since
$p_{R}$ corresponds to the momentum of $\alpha$ in the initial state
in quasifree knockout reactions.

\begin{figure}[bp]
\centering
\includegraphics[width=0.42\textwidth]{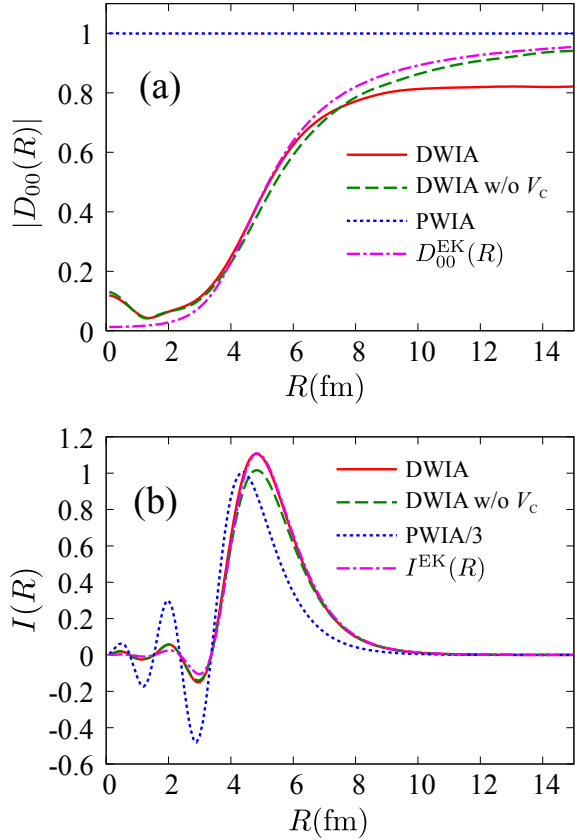}
\caption{
(a)
Masking function
$\left| D_{00}(R) \right|$ defined by
Eq.~(\ref{eqmask}).
The solid, dashed, and dotted lines represent
$\left| D_{00}(R) \right|$
of DWIA, DWIA without Coulomb interactions, and PWIA, respectively.
The dot-dashed line shows the eikonal-masking function
$\left| D_{00}^\mathrm{EK}(R) \right|$ discussed in Sec.~\ref{subsecEK}.
(b)
Same as (a) but $I(R)$ and that with the
eikonal approximation $I^\mathrm{EK}(R)$.
}
\label{figDRandIR}
\end{figure}

In Fig.~\ref{figDRandIR} we investigate the
masking function $D_{00}(R)$
and the radial reaction amplitude $I(R)$ at $p_{R}=0$~MeV/$c$. The latter is defined by
\begin{align}
I(R)
&\equiv
R^2 \left|D_{00}(R)\right| \phi_{\alpha}^{40}(R),
\label{eqi}
\end{align}
which corresponds to the integrand of $R$ in Eq.~(\ref{eqmask}).
As shown by the solid line in Fig.~\ref{figDRandIR}(a),
$\left|D_{00}(R)\right|$ is enough small in the nuclear interior region to suppress
contribution of the $\alpha$ amplitude in this region to $I(R)$ because of
the absorption effect of the distorting potentials.
The result without the Coulomb interaction $V_c$
(dashed line in Fig.~\ref{figDRandIR}(a))
approaches to unity in the asymptotic region because
all the nuclear potentials are absent there and
$F_{{\bm K}_i}({\bm R})=1$
when $q=0$; see Eqs.~(\ref{eqF}) and (\ref{eqmomtrans}).
On the other hand, the result of DWIA including $V_c$ (solid line)
never reaches to unity even in the asymptotic region.
This is because $F_{\bm{K}_i}({\bm R})$ suffers from
the long-range nature of the Coulomb interaction.
Even if the asymptotic
momenta of the three particles satisfy $\bm{q}=0$,
finite values of the Coulomb phase shifts of the scattering waves
make $\left|D_{00}(R)\right|$ deviate from unity when $R$ is finite.

Figure~\ref{figDRandIR}(b) shows $I(R)$ defined by Eq.~(\ref{eqi}).
One sees that $I(R)$ of DWIA (solid line) is strongly suppressed in the
interior region, compared with that of PWIA divided by 3 (dotted line).
The result of DWIA without Coulomb interactions (dashed line) agrees well
with that of DWIA (solid line),
since the cluster wave function has an amplitude for $R\la8$~fm,
where $\left|D_{00}(R)\right|$ with and without Coulomb interactions
agrees well each other.
Therefore the deviation of the solid line
in Fig.~\ref{figDRandIR}(a) from unity due to Coulomb interactions
makes no difference in understanding the property of $D_{00}(R)$.
This allows one to make further simplification of $D_{00}(R)$
as discussed in Sec.~\ref{subsecEK}.

The masking effect on the cluster wave function due to
nuclear distorting potentials can be
a great advantage of
($p$, $p\alpha$) reactions for probing $\alpha$ cluster states.
A large ($p$, $p\alpha$) cross section
corresponds to a large cluster amplitude around the nuclear surface,
which indicates an existence of well-developed $\alpha$ cluster states in a nucleus.
It should be noted that in the
nuclear interior, the NN antisymmetrization plays a significant role,
which makes even the definition of the $\alpha$ cluster unclear. Because of
the masking effect, ($p$, $p\alpha$) reactions automatically avoid such an
unclear region and focus the nuclear surface. In consequence of this,
one can conclude that what is determined by ($p$, $p\alpha$) is not an
$\alpha$ spectroscopic factor $S_{\alpha}$ but the surface amplitude of the $\alpha$
distribution, and the latter is the measure of the $\alpha$ clustering.

\subsection{Coulomb effect in $^{120}$Sn($p$, $p\alpha$)$^{116}$Cd reaction}
As shown in Fig.~\ref{figDRandIR} (a),
long-ranged
Coulomb interactions
prohibit the masking function from reaching unity,
even if the kinematics of the knockout process is fixed so as
to satisfy $\bm{q}=0$.
This effect becomes significant when the charge of the target nucleus
is large.
In this point of view,
we investigate the peripheral property of the
$^{120}$Sn($p$, $p\alpha$)$^{116}$Cd reaction at
392~MeV and the masking function as well.
The $\alpha$ cluster wave function is constructed in
the same method as mentioned above but with
$r_0=1.25\times 116^{1/3}$~fm and
$\varepsilon_{\alpha}=-4.81$~MeV.
The three-body kinematics are chosen to be
$T_1=328$~MeV, $\theta_1^\mathrm{L} = 43.2^\circ$,
$T_\alpha=51$~MeV, and $\theta_\alpha=61^\circ$,
which satisfies $\bm{q}=0$.

\begin{figure}[htbp]
\centering
\includegraphics[width=0.45\textwidth]{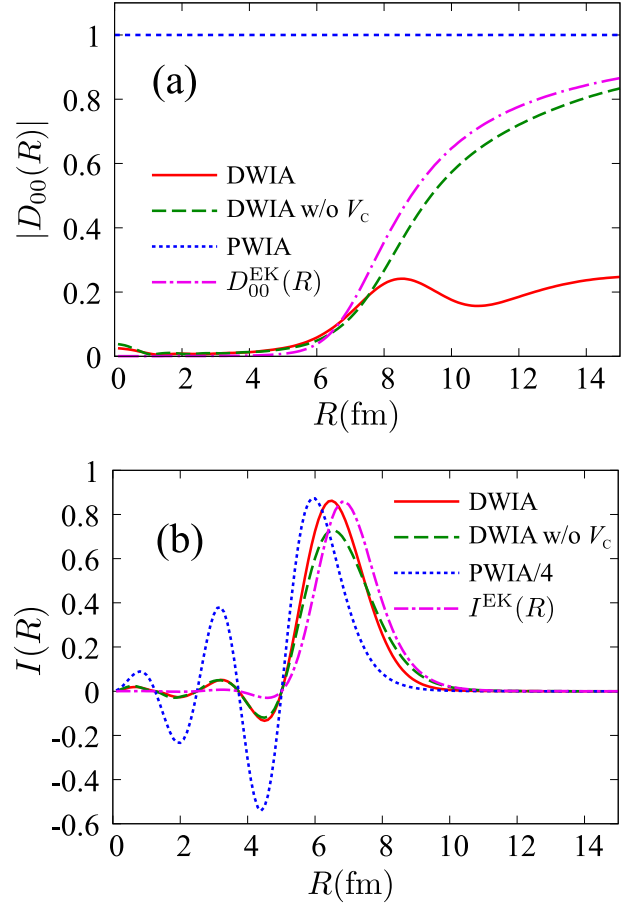}
\caption{
The same as Fig.~\ref{figDRandIR} but for $^{120}$Sn($p$, $p\alpha$)$^{116}$Cd
reaction.
}
\label{figDRIRSn}
\end{figure}

In Fig.~\ref{figDRIRSn} $\left|D_{00}(R)\right|$ and $I(R)$ of
$^{120}$Sn($p$, $p\alpha$)$^{116}$Cd are shown in the
same manner as in Fig.~\ref{figDRandIR}.
The obtained masking function for this system
(the solid line in Fig.~\ref{figDRIRSn} (a))
is only
$0.2$--$0.3$ even
in the asymptotic region
by the Coulomb interactions.
On the other hand, $D_{00}(R)$ without
$V_C$ (the dashed line)
tends to approach unity, as well as
the eikonal masking function $D_{00}^{\mathrm{EK}}(R)$
(dot-dashed line).
This peculiar behavior of $D_{00}(R)$ at larger $R$, however,
causes no serious problem, because the $\alpha$ cluster wave function
is sufficiently damped in that region.
Consequently, the dashed line
and the dot-dashed line agree quite well with the solid line in
Fig.~\ref{figDRIRSn} (b)
indicating that $I(R)$ can be approximately simulated by analyses of the
masking function without Coulomb interactions and
also the eikonal masking function.
It turns out that only the $\alpha$ amplitude at the nuclear surface can be safely probed by $\alpha$ knockout reactions
even in the case of heavy-mass targets.

\subsection{Incident energy dependence of masking function}
\label{subsecEK}
It is important to clarify how the aforementioned property of the
masking function depends on the incident energy.
For this purpose, a more simplified functional form of the
masking function is preferable. We thus rely on the eikonal
approximation and use the following form:
\begin{align}
D_{00}^{\mathrm{EK}}(R)
&\equiv
\frac{1}{4\pi}
\int d\Omega\,
\mathrm{exp}
\left[
-C_{\mathrm{abs}}
\int_{-\infty}^{\infty} dz
f_{\mathrm{WS}}({\bm R})
\right]
e^{-i{\bm q} \cdot {\bm R}},
\label{eqekmask}
\end{align}
which is designated eikonal masking function.
$C_{\mathrm{abs}}$ represents
the total strength of the absorption caused by the distorting potentials
of particles 0--2, and is determined so that the peak height of
\begin{align}
I^{\mathrm{EK}}(R)
&\equiv
R^2
\left|
D_{00}^{\mathrm{EK}}(R)
\right|
\phi_{\alpha}^{40}(R)
\end{align}
reproduces that of $I(R)$.
Thus, one can characterize the masking function $D_{00}(R)$ by
just one parameter $C_{\mathrm{abs}}$.
In a na\"ive interpretation, $C_\mathrm{abs}$ is related to
the mean free path (MFP) $\lambda$ through
$1/\lambda = 2C_\mathrm{abs}f_\mathrm{WS}(R)$.

\begin{figure}[htbp]
\centering
\includegraphics[width=0.45\textwidth]{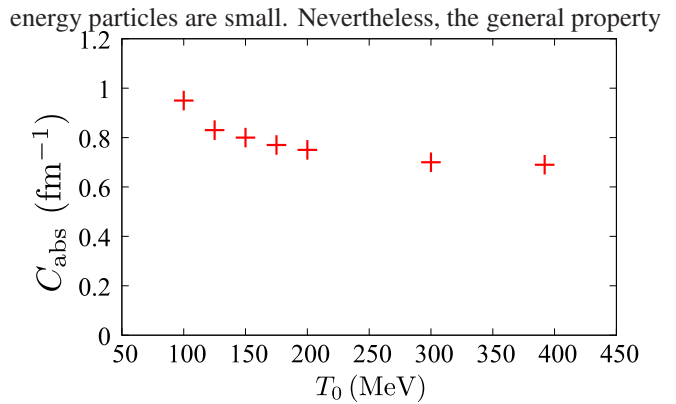}
\caption{
Incident energy $T_0$ dependence of $C_\mathrm{abs}$.
}
\label{figcfac}
\end{figure}

The dot-dashed lines in Fig.~\ref{figDRandIR}(a) and \ref{figDRandIR}(b) represent
$\left|D_{00}^{\mathrm{EK}}(R)\right|$ and $I^{\mathrm{EK}}(R)$,
respectively; $C_\mathrm{abs}$ is taken to be $0.69$~fm$^{-1}$.
One sees in Fig.~\ref{figDRandIR}(a) that the dot-dashed line
agrees well with the solid line for $4 \la R\la 6$~fm and slightly
deviates from it for larger $R$. This asymptotics of the dot-dashed
line can be understood as a result of absence of Coulomb interactions
in Eq.~(\ref{eqekmask}).
Nevertheless, as shown in Fig.~\ref{figDRandIR}(b), the dot-dashed line,
$I^\mathrm{EK}(R)$,
agrees well with the solid line for $R\ga 4$~fm, in which
the integrand $I(R) $ has a meaningful amplitude. This indicates
that Eq.~(\ref{eqekmask}) is sufficient
to describe the property of the masking function that is relevant to the
($p,p\alpha$) reaction considered. It should be noted that
the small but finite difference between the solid and dot-dashed lines in
Fig.~\ref{figDRandIR}(a) for $R\ga 6$~fm appears also in
Fig.~\ref{figDRandIR}(b); for instance, the difference is about 6\% at 8~fm.
We then discuss the incident energy $T_0$ dependence of $C_{\mathrm{abs}}$,
which is shown in Fig.~\ref{figcfac}.
At each $T_0$, the scattering energies and angles of particle 1 and 
$\alpha$ are chosen so as to satisfy $p_R = 0$~MeV/$c$.
It is found that the $T_0$ dependence of $C_\mathrm{abs}$ is
weak above $200$~MeV, where the quasi-free (impulse) picture
of the knockout reaction is valid.
On the other hand, below $200$~MeV, $C_\mathrm{abs}$ increases rather
rapidly as $T_0$ decreases. This behavior can be understood
by the fact that the MFP's of low energy particles are small.
Nevertheless, the general property of the masking function
and the peripherality of the reaction are
found to be robust for $T_0$ shown in Fig.~\ref{figcfac}.

\section{Summary}
\label{secsum}
$^{20}$Ne($p$, $p\alpha$)$^{16}$O reactions at 100--392~MeV were investigated
within the distorted wave impulse approximation framework.
We have introduced the masking function, which describes the absorption
effect due to distorting potentials of the incident $p$ and also the
emitted $p$ and $\alpha$ and suppresses the $\alpha$ amplitude
contribution in the interior region to the
$\alpha$ knock-out  cross section.
Through the analyses on the masking functions of the reactions,
it is clearly shown that $\alpha$ knockout reactions are peripheral
and suitable for probing the $\alpha$ cluster amplitude in the
nuclear surface, which is regarded as a direct
measure of spatially
developed $\alpha$ cluster states.

As a case of strong Coulomb interactions,
the $^{120}$Sn($p$, $p\alpha$)$^{116}$Cd reaction at 392~MeV
has been investigated.
It is shown that
in this case the masking function has a nontrivial behavior
at larger distance
due to Coulomb interactions.
Nevertheless, this
does not cause any problem because the overlap between the
masking function and the cluster amplitude in that region is
negligibly small.

To investigate the incident energy $T_0$ dependence of the masking
function, we introduce a simplified functional that is
characterized by only one parameter $C_\mathrm{abs}$.
It was found that $C_\mathrm{abs}$ depends weakly on $T_0$
for 200--400~MeV. At lower energies, $C_\mathrm{abs}$ increases
rather rapidly, corresponding to stronger absorption.
However, the feature of the masking function turned out to be
robust down to 100~MeV.

\section*{ACKNOWLEDGMENTS}
A part of the computation
was carried out with the computer facilities at the Research
Center for Nuclear Physics, Osaka University.
This work was
supported in part by Grants-in-Aid of the Japan Society for
the Promotion of Science (Grants No. JP16K05352,
JP15J01392, and No. JP26400270).


\end{document}